\documentclass{ws-ijmpa}
\usepackage{graphicx}

\begin{document}

\markboth{Kaplan}{A New Experiment to Study Hyperon CP Violation...}

\catchline{}{}{}{}{}

\title{A NEW EXPERIMENT TO STUDY HYPERON CP VIOLATION AND THE CHARMONIUM SYSTEM\footnote{To appear in Proceedings of CTP symposium on Supersymmetry at LHC: Theoretical and Experimental Perspectives, The British University in Egypt, Cairo, Egypt, 11--14 March 2007. } 
}
\author{DANIEL M. KAPLAN}
\address{Physics Division, Illinois Institute of Technology\\
 Chicago, Illinois 60616, USA\\
kaplan@iit.edu}

\maketitle

\begin{history}
\received{15 June 2007}

\end{history}

\begin{abstract}
Fermilab operates  the world's most intense antiproton source, now exclusively dedicated to serving the needs of the Tevatron Collider. 
The anticipated 2009 shutdown of the Tevatron presents the opportunity for a world-leading low- and medium-energy antiproton program. We summarize the  status of the Fermilab antiproton facility and review physics topics for which a future experiment could make the world's best measurements.

\end{abstract}

\section{Introduction}

An international collaboration is proposing\cite{pbar-LoI} to revive the experimental program at the Fermilab Antiproton Accumulator. We summarize the goals of this effort in the context of other efforts world-wide.

\subsection{Antiproton sources}
The world's highest-energy and highest-intensity antiproton source is at Fermilab. Having previously supported medium-energy antiproton fixed-target experiments (including the charmonium experiments E760 and E835), it is now 100\% dedicated to providing luminosity for the Tevatron Collider. At CERN, the LEAR antiproton storage ring was decommissioned in 1996;
its successor facility, the Antiproton Decelerator (AD), provides antiproton beams at  momenta of 100 and 300\,MeV/$c$, at intensities up to $\approx2\times10^7$ per minute.\cite{Eriksson}
These are the only operating facilties. 
Germany has embarked on  a $\approx$billion-Euro upgrade  for the GSI-Darmstadt nuclear-physics laboratory, with planned construction by $\approx$\,2014 of 30 and 90\,GeV rapid-cycling synchrotrons and low- and medium-energy antiproton storage rings.\cite{FAIR}

\subsection{Physics with antiproton sources}\label{physics-list}
Many interesting topics can be addressed with such a facility, including 
\begin{itemize}
\item precision $\overline{p}p\to\,$charmonium studies, begun by Fermilab E760 and E835;

\item open-charm studies, including searches for $D^0/{\overline D}{}^0$ mixing and {\em CP} violation;

\item
studies of  $\overline{p}p\to\,$hyperons, including 
hyperon {\em CP} violation and rare decays;

\item
the search for glueballs and gluonic hybrid states predicted by QCD;
and

\item
trapped-$\overline{p}$ and antihydrogen studies.

\end{itemize}

\noindent
Needed beam energy or intensity makes only the last of these possible at the CERN AD. All have been discussed as program components of the GSI-FAIR (Facility for Antiproton and Ion Research) project\cite{FAIR} and its general-purpose PANDA detector.\cite{PANDA-TPR} However,  GSI-FAIR construction has yet to begin, and PANDA data taking is not expected before 2014. Table~\ref{tab:thresh} gives some relevant mass and momentum thresholds.

\begin{table}[h]
\tbl{Thresholds for some processes of interest in $\sqrt{s}$ and $\overline{p}$ momentum in the lab frame for  $\overline{p}p$ fixed-target.}
{\begin{tabular}{@{}lcc@{}} \toprule
 & \multicolumn{2}{c}{Threshold} \\
Process & $\sqrt{s}$ & $p_{\overline p}$\\
 & (GeV) &  (GeV/$c$) \\
\colrule 
$\overline{p}p\to\overline{\Lambda}\Lambda$  & 2.231 & 1.437\\
$\overline{p}p\to\overline{\Sigma}{}^-\Sigma^+$  & 2.379 & 1.854\\
$\overline{p}p\to\overline{\Xi}{}^+\Xi^-$ & 2.642 & 2.620 \\
$\overline{p}p\to\overline{\Omega}{}^+\Omega^-$ & 3.345 & 4.938\\
\colrule 
$\overline{p}p\to\eta_c$ & 2.980 & 3.678\\
$\overline{p}p\to\psi(3770)$ & 3.771 & 6.572  \\
$\overline{p}p\to X(3872)$ & 3.871 & 6.991  \\
$\overline{p}p\to X {\rm \,or\,} Y(3940)$ & 3.940 & 7.277  \\
$\overline{p}p\to Y(4260)$ & 4.260 & 8.685  \\
\botrule 
\end{tabular}
\label{tab:thresh}}
\end{table}

A number of intriguing recent discoveries  can be elucidated at such a facility: the states provisionally named $X(3872)$, $X(3940)$, $Y(3940)$, $Y(4260)$, and $Z(3930)$ in the charmonium region,\cite{ELQ} as well as the observation of apparent flavor-changing neutral currents (FCNC) in hyperon decay;\cite{Park-Sigpmumu} indeed, high sensitivity can be achieved to symmetry-violating or rare hyperon decays generally. In addition, the $h_c$ mass and width, $\chi_c$ radiative-decay angular distributions, and ${\eta^\prime_c}(2S)$ full and radiative widths, important parameters of the charmonium system that remain to be precisely determined, are well suited to the ${\overline p}p$ technique.\cite{E835-NIM,E835-psi-widths}

\subsection{\it Quarkonium physics} 
\begin{figure}
\vspace{-1.1in}
\centerline{\includegraphics[width=.9\linewidth]{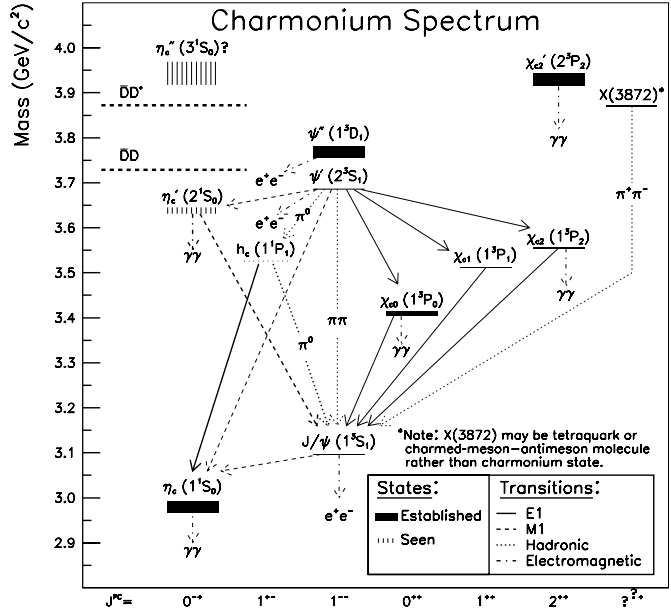}}
\vspace{-1.6in}
\caption{Spectrum of the charmonium system. Shown are masses, widths (or for those not yet measured, 90\% confidence level upper limits on widths), and quantum numbers of observed charmonium states, with some of the important transitions also indicated.\protect\cite{PDG2006,ELQ}}
\label{fig:ccbar}
\end{figure}

Heavy-quark--antiquark bound states (``quarkonia") offer a unique testing ground for QCD. Both potential models and lattice-gauge Monte Carlo successfully predict aspects of heavy-quark systems.
Quenched-approximation lattice-QCD predictions of the masses of low-lying charmonium states agree qualitatively with the experimental values;\cite{QWG-Yellow} this agreement is expected to improve once dynamical quarks on the lattice (now being implemented by various groups) are successfully incorporated.\cite{LQCD} Charmonium (Fig.~\ref{fig:ccbar}) is an important proving ground for QCD: the $c$ and $\overline c$ quarks are slow enough that relativistic effects are significant but not dominant, and are sufficiently massive that non-perturbative effects are small but not negligible. Once certified by experiment, these calculational techniques can then be confidently applied in interpreting such physics results as {\em CP} asymmetries in the beauty system.

Fermilab experiments E760 and E835 made the world's most precise measurements of charmonium masses and widths.\cite{E835-NIM,E835-psi-widths} This precision ($\stackrel{<}{_\sim}$\,100\,keV) reflects the narrow energy spread of the stochastically cooled antiproton beam and the absence of Fermi motion and negligible energy loss in hydrogen cluster-jet targets. The other key advantage of ${\overline p}p$  annihilation is its ability to produce charmonium states of all quantum numbers, whereas $e^+e^-$ machines produce primarily $1^{--}$ states. 

\subsection{\it Our proposal} We are proposing a focused experimental program aimed at those measurements for which the Antiproton Source is best suited: (1) precision studies of states in the charmonium region and (2) the search for new physics in hyperon decay. These measurements can 
be performed with a common apparatus using existing technologies. Depending on available resources, existing detector components might be recycled for these purposes; alternatively, modest expenditures for new equipment could yield improved performance. The opportunity for such studies will soon arrive, with the planned 2009 shutdown of the Tevatron.  The importance of these measurements justifies the resumption of such a program at Fermilab.

\section{Capabilities of the Fermilab Antiproton Source}

The Antiproton Source maximum stacking rate
is now $\approx$\,20\,mA/hr (or $2\times10^{11}\,{\overline p}$/hr), five times that in E835.\cite{E835-NIM} We propose to run with up to ten times the typical E835 luminosity\cite{E835-NIM} (${\cal L}\stackrel{<}{_\sim}2\times10^{32}\,{\rm cm}^{-2}{\rm s}^{-1}$), via increased store intensity or  target density. Since stochastic cooling works best with small stacks, more intense stores seem nonoptimal. 
The E835 cluster-jet target (an upgrade of the E760 one) produced\cite{E835-NIM} up to $\approx2.5\times10^{14}\,$atoms/cm$^2$. Higher cluster-jet density is proposed for the PANDA program (also planned for ${\cal L}=2\times10^{32}\,{\rm cm}^{-2}{\rm s}^{-1}$).\cite{PANDA-TPR}
Other options include a plastic or metal wire or pellet in the beam halo,\cite{HERA-B} a solid-H$_2$ target on the tip of a cold finger, or a stream of H$_2$ pellets. 
A non-H$_2$ target, while suitable for hyperon running,  would destroy the superb energy resolution needed for the charmonium studies. We favor simultaneous charmonium and hyperon running with an H$_2$ target.

\section{Physics Goals}

To clarify issues for a future antiproton facility,
we consider  representative physics examples: studying the $X(3872)$, improved measurement of the parameters of the $h_c$, searching for hyperon $C\!P$ violation, and studying a recently discovered rare hyperon-decay mode. (This list is not exhaustive; see Sec.~\ref{addl-phys} for additional topics.)

\subsection{$X(3872)$}
The $X(3872)$ was discovered\cite{Belle-3872} in 2003 by the Belle Collaboration via $B^\pm\to K^\pm X(3872)$, $X(3872)\to \pi^+\pi^-J/\psi$, and quickly confirmed by CDF,\cite{CDF-3872} D\O,\cite{D0-3872} and BaBar.\cite{BaBar-3872} Now seen (Table~\ref{tab:X3872}) in  $\gamma J/\psi$,\cite{Belle-Jpsi-gamma} $\pi^+\pi^-\pi^0J/\psi$,\cite{Belle-3piJ} and $D^0{\overline D}{}^0\pi^0$ modes\cite{Belle-DDpi} as well, it does not appear to fit within the charmonium spectrum. Although well above open-charm threshold, its  width\cite{PDG2006} ($<2.3$\,MeV at 90\% C.L.) implies that decays to $D{\overline D}$ are forbidden, suggesting unnatural parity,\cite{QWG-Yellow} $P=(-1)^{J+1}$. It is a poor candidate for  $\psi_2\,(1\,^3D_2)$ or $\psi_3\,(1\,^3D_3)$\cite{ELQ,Belle-3piJ,QWG-Yellow} due to nonobservation of radiative transitions to $\chi_c$. The observation of $X (3872) \to \gamma J/\psi$ implies positive $C$-parity, and additional observations essentially rule out all possibilities other than $J^{PC}=1^{++}$.\cite{Belle-LP2005,Braaten-HQW} The  available charmonium assignment with those quantum numbers,  $\chi^\prime_{c1}\,(2\,^3P_1)$, is highly disfavored\cite{ELQ,QWG-Yellow} by the observed rate of $X (3872) \to \gamma J/\psi$. Moreover, the plausible identification of $Z(3930)$ as the $\chi^\prime_{c2}\,(2\,^3P_2)$  suggests\cite{ELQ} that the $2\,^3P_1$  should lie some 49\,MeV/$c^2$ higher than the observed\cite{PDG2006} $m_X=3871.2\pm0.5\,$MeV/$c^2$.

\begin{table}[h]
\tbl{Experimental observations of $X(3872)$.}
{\begin{tabular}{@{}lccccc@{}} \toprule
Experiment & Year & Mode & Events & Ref.\\
\colrule
Belle & 2003 & $\pi^+\pi^-J/\psi$ & $35.7\pm6.8$ & \refcite{Belle-3872}\\
CDF & 2004 & $\pi^+\pi^-J/\psi$ & $730\pm90$ & \refcite{CDF-3872}\\
D0 & 2004 &$\pi^+\pi^-J/\psi$ & $522\pm100$ &  \refcite{D0-3872} \\
Belle  & 2004 &$\omega (\pi^+\pi^-\pi^0)J/\psi$ & $10.6\pm3.6$& \refcite{Belle-3piJ} \\
BaBar & 2005  &  $\pi^+\pi^-J/\psi$ &$25.4\pm8.7$  & \refcite{BaBar-3872}\\
Belle  & 2005 &$\gamma J/\psi$ & $13.6\pm4.4$ &\refcite{Belle-Jpsi-gamma} \\
Belle & 2006&  $D^0{\overline D}{}^{*0}$ & $23.4\pm5.6$ & \refcite{Belle-DDpi}\\
\botrule
\end{tabular}\label{tab:X3872}}
\end{table}

The coincidence of the $X(3872)$ with  $D^0 {\overline D}{}^{*0}$  threshold suggests various 
solutions to this puzzle, including an $S$-wave cusp\cite{Bugg} or a tetraquark state.\cite{tetraquark} An intriguing possibility is that the $X(3872)$ represents the first clear-cut observation of a meson-antimeson molecule: a bound state of $D^0 {\overline D}{}^{*0}+D^{*0} {\overline D}{}^0$.\cite{molecule}\,\footnote{The mass coincidence may be accidental, and the $X(3872)$ a $c{\bar c}$-gluon hybrid state; however, the mass and 1$^{++}$ quantum numbers make it a poor match to lattice-QCD predictions for such states.\protect\cite{ELQ}} A key measurement is the precise mass difference between the $X$ and that threshold, which should be  slightly negative, in accord with the small molecular binding energy:\cite{Braaten-HQW}
\begin{equation}0<E_X=(m_{D^0}+m_{D^{*0}}-m_X)c^2\ll10\,{\rm MeV}\,.\end{equation}
A measurement of the width is also highly desirable. 

With the latest CLEO measurement,\cite{CLEO-mD} $M_{D^0}= 1864.847 \pm 0.150 \pm 0.095$\,MeV/$c^2$, and the world-average\cite{PDG2006} $m_{D^{*0}}-m_{D^0}=142.12\pm0.07$\,MeV/$c^2$, we have $E_X=0.6\pm0.6$\,MeV/$c^2$, with the uncertainty dominated by that of $m_X$. When our precision measurement is made, it will still  dominate, assuming the total uncertainty on $m_{D^0}$ improves roughly as $1/\sqrt{N}$ as the statistics of the CLEO analyzed sample increase by an order of magnitude.\cite{Rosner-private} Additional important measurements include ${\cal B}[X(3872)\to \pi^0\pi^0J/\psi]$ to confirm the $C$-parity assignment\cite{Barnes-Godfrey} and ${\cal B}[X(3872)\to \gamma\psi^\prime]$ to further tighten the constraints with respect to the $2\,^3P_1$ assignment.\cite{ELQ}

\subsubsection{$X(3872)$ sensitivity estimate}

The ${\overline p}p\to X(3872)$ cross section  is unmeasured but  estimated to be similar in magnitude to those for $\chi_c$.\cite{Braaten-X-production} This estimate is supported by the observed rates and  distributions of ${\overline p}p\to X(3872)\,+$\,anything at the Tevatron\cite{D0-3872} and of $B^\pm\to K^\pm X(3872)$,\cite{PDG2006} which resemble those for charmonium states. E760 detected $\chi_{c1},\, \chi_{c2}\to\gamma J/\psi$ 
(branching ratios of 36\% and 20\%, respectively\cite{PDG2006}) with  $44\pm2$\% acceptance $\times$ efficiency and $\approx$\,500 observed events  per ${\rm pb}^{-1}$ at each resonance.\cite{E760-chi_c} At $10^{32}\,{\rm cm}^{-2}{\rm s}^{-1}$, the 90\%-C.L.  limit\cite{BaBar-BR} ${\cal B}[X(3872)\to\pi^+\pi^-J/\psi]>0.042$ then 
implies $\stackrel{>}{_\sim}$\,4\,$\times10^3$ events in that mode per nominal month ($1.0\times10^6$\,s) of running. Current sample sizes (Table~\ref{tab:X3872}) are likely to increase by not much more than an order of magnitude as  experiments complete during the current decade.\footnote{The ${\overline p}p\to X(3872)$ sensitivity will be competitive even with that of the proposed SuperKEKB upgrade,\protect\cite{SuperKEKB} should that project go forward.}

Given the uncertainties in the cross section and branching ratios, the above may well be an under- or overestimate of the ${\overline p}{p}$ formation and observation rates, perhaps by as much as an order of magnitude. Nevertheless, it appears that a new experiment at the Antiproton Accumulator could obtain the world's largest clean samples of $X(3872)$, in perhaps as little as a month of running. The  high statistics, event cleanliness, and unique precision available in the ${\overline p}p$ formation technique could enable the world's smallest systematics. Such an experiment could thus provide a definitive test of the nature of the $X(3872)$.

\subsection{\boldmath $h_c$}

Observing the $h_c$\,$(1 {}^1P_1)$ charmonium state and measuring its parameters were high-priority goals of E760, E835, and their predecessor experiment, CERN R704. As a narrow state with suppressed couplings both to $e^+e^-$ and to the states that are easily produced in $e^+e^-$ annihilation, the $h_c$ is a difficult state to study experimentally.

The pioneering charmonium experiment R704 was one of the last experiments at the Intersecting Storage Rings (ISR).  With a stochastically cooled $\overline p$ beam, a hydrogen 
cluster-jet target, and a nonmagnetic spectrometer of limited angular coverage, it searched for final states including $J/\psi\to e^+e^-$ decay. A claimed 2.3$\sigma$ signal of 5 ${\overline p}p\to J/\psi + X$ events near the $\chi_c$ center of gravity was interpreted as evidence for the (isospin-violating) $h_c\to J/\psi\,\pi^0$ mode, with $h_c$ mass $3525.4\pm0.8\pm0.5$\,MeV.\cite{Baglin} The R704 signal implies an on-resonance cross section of $\approx$\,2\,nb and $\Gamma_{h_c} \times {\cal B}(h_c\to p{\overline p}) \times {\cal B}(h_c\to J/\psi\, X) \times {\cal B}(J/\psi\to e^+e^-)=  0.135^{+0.150}_{-0.060}$\,eV.\cite{Cester-Rapidis}

Following the ISR shutdown, these studies continued with Fermilab E760. Also nonmagnetic but with $\approx$\,3$\pi$ coverage, E760 found in a $\approx$\,17\,pb$^{-1}$ sample an enhancement in $J/\psi\,\pi^0$ at $3526.2 \pm0.15\pm0.2$\,MeV,\cite{E760-h_c} with a 1-in-400 estimated probability of  the 59 candidate events arising at random. The E760 measurements of on-resonance cross-section, $\approx$\,0.3\,nb, and $\Gamma_{h_c} \times {\cal B}(h_c\to p{\overline p}) \times {\cal B}(h_c\to J/\psi\, X) \times {\cal B}(J/\psi\to e^+e^-)=  0.010 \pm 0.003$\,eV  appear to rule out the R704 events as being signal.\cite{Cester-Rapidis}

E835 spent considerable running time in the $h_c$ region, finding no signal in $J/\psi\,\pi^0$ at the E760 mass value, but  (in an $\approx$\,80\,pb$^{-1}$ sample)   a 13-event enhancement in $\eta_c\gamma$ (with  $\eta_c\to\gamma\gamma$) at $3525.8 \pm0.2\pm0.2\,$MeV.\cite{E835-h_c} The estimated significance, in the range 1--$3\times10^{-3}$, was comparable to that of the E760 $h_c$ signal. More recently, CLEO's study\cite{CLEO-h_c} of $\psi(2S)\to\pi^0 h_c \to(\gamma\gamma)(\gamma\eta_c)$ has  established the existence of the $h_c$ at $>$\,4$\sigma$: based on $168\pm40$ signal events, they find $m(h_c)=3524.4\pm0.6\pm0.4$\,MeV, not inconsistent with the  E835 measurement. Neither experiment was able to measure the width of the $h_c$, but E835 set a 90\%-C.L. upper limit of 1\,MeV. 

\begin{figure}
\vspace{-.025in}
\centerline{\includegraphics[width=3 in]{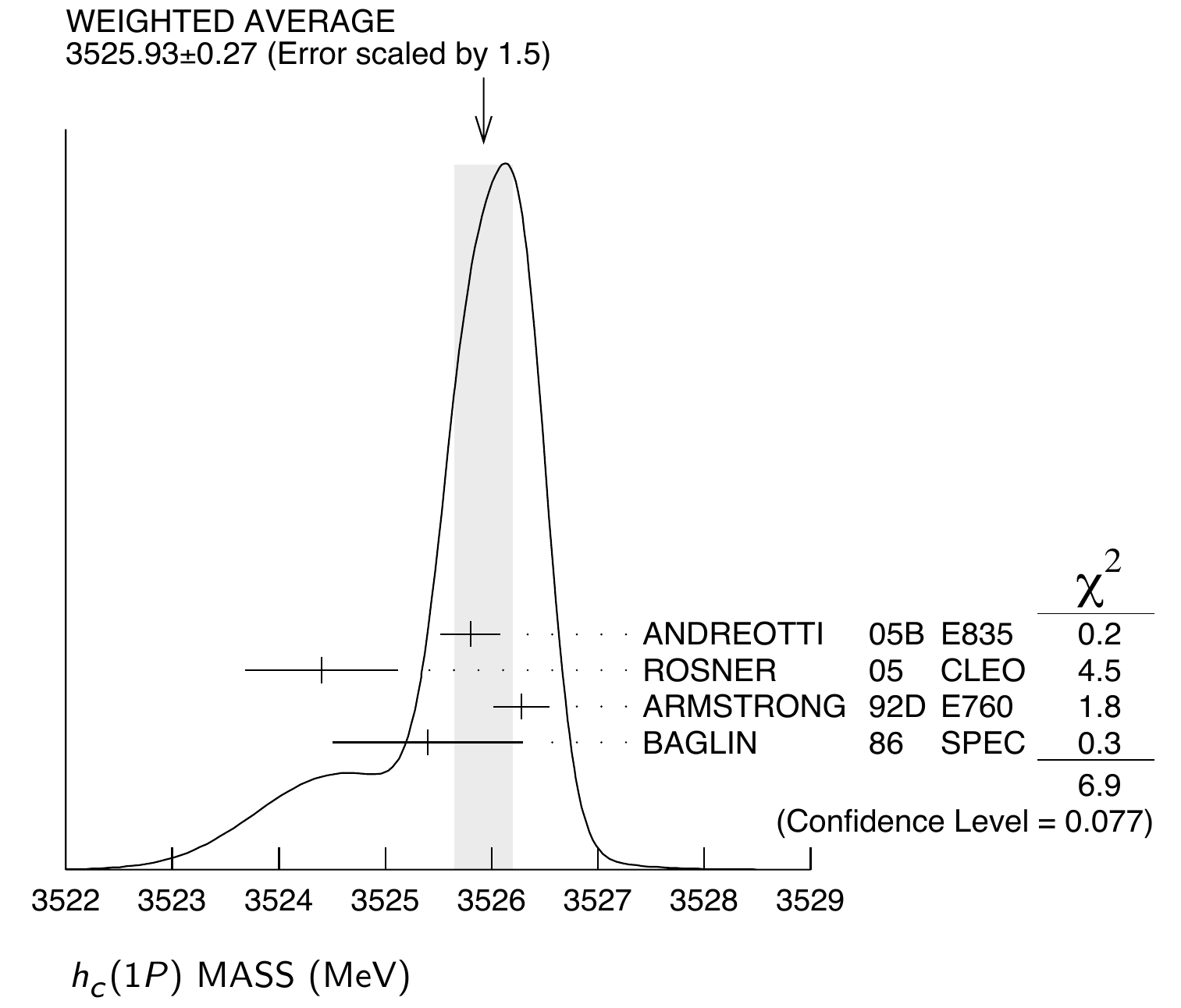}}
\vspace{-.1in}
\caption{PDG ideogram of the four most precise measurements of the $h_c$ mass (from Ref.~\protect\refcite{PDG2006}).}\label{fig:hc-ideogram}
\end{figure}

A key prediction of QCD and perturbation theory is that the charmonium spin-zero hyperfine splitting, as measured by the mass difference $\Delta m_{\rm hf}$ between the $h_c$ and the spin-weighted average of the $\chi_c$ states, should be close to zero.\cite{HF-pred} Using the current PDG-average values,\cite{PDG2006} $\langle m(^3P_J)\rangle = 3525.36\pm0.06$\,MeV and $m(h_c)=3525.93\pm0.27$\,MeV, we find $\Delta m_{\rm hf}=-0.57\pm0.28$\,MeV, nonzero at 2$\sigma$ but within the QCD expected range. The PDG error on $m(h_c)$ includes a scale factor of 1.5 due to the tension among the four most precise measurements (Fig.~\ref{fig:hc-ideogram}). Moreover, the two most precise (E760 and E835) are based on statistically marginal signals, and the reliability of the  E760 result is called into question by the negative results of the E835 search. The R704 result is on even weaker ground: a ${\overline p}p\to h_c\to J/\psi\, X$ signal at the level implied by Baglin {\it et al.}\cite{Baglin} is most likely ruled out by  E760\cite{Cester-Rapidis} (as discussed above) as well as by E835. Thus of the four results used by the PDG in Fig.~\ref{fig:hc-ideogram}, only one is clearly reliable, and the claimed precision on $m(h_c)$ is far from established. This motivates an improved experimental search. Also of  interest are the width and branching ratios of the $h_c$, for which QCD makes clear predictions; the decay modes also bear on the question of isospin conservation in such decays.

E835's  $h_c\to\eta_c\gamma\to(\gamma\gamma)\gamma$ sensitivity was limited by the $(2.8\pm0.9)\times10^{-4}$  $\eta_c\to\gamma\gamma$ branching ratio, and their acceptance $\times$ efficiency was  only $\approx$\,3\% due to cuts against the substantial $\pi^0$ background.\cite{E835-h_c} With a magnetic spectrometer, likely $\eta_c$ modes include $\phi\phi$, $\phi K^+K^-$, $K^*K^*$,  and $\eta^\prime\pi^+\pi^-$. These have branching ratios up to two orders of magnitude larger, as well as more-distinctive decay kinematics, than $\gamma\gamma$, probably allowing looser cuts and thus higher efficiency. For example, the $\phi\phi\to K^+K^-K^+K^-$ final state has no quarks in common with the initial ${\overline p}p$ state and so should contain little background. E835 searched for $\eta_c\to\phi\phi$  but without a magnet it was barely feasible. Assessing the degree of improvement will require detailed simulation work, but at least an order of magnitude in statistics seems likely. Additional improvement will come from the higher luminosity we propose.

Provided detailed simulation studies bear out these ideas, we will soon have the opportunity to resolve this  20-year-old experimental controversy.

\subsection{Hyperon {\boldmath $C\!P$} violation}

Besides the well-known  {\em CP} violation in $K$- and $B$-meson mixing and decay,\cite{PDG2006} the standard model (SM) predicts slight hyperon-decay asymmetries.\cite{Hyperon-CP}$\!^{-\!}$\cite{Valencia2000} Standard-model processes dominate $K$ and $B$ {\em CP} asymmetries, thus it behooves us to study  hyperons (and charm; Sec.~\ref{addl-phys}), in which new physics might stand out more sharply. 

More than one hyperon $C\!P$ asymmetry may be measurable in ${\overline p}p$ annihilation. Hyperon {\em CP} violation would of course be of the direct type, so as to conserve baryon number. Accessible signals include angular-distribution differences of polarized-hyperon and antihyperon  decay products;\cite{ACP} partial-rate asymmetries, possibly at detectable levels, are also expected.\cite{Tandean-Valencia,Tandean} To  compete with previous $\Xi$ and $\Lambda$ {\em CP} studies would require $\sim10^{33}$ luminosity. While summarizing the state of hyperon {\em CP} violation generally, we therefore emphasize in particular the $\Omega^-/{\overline \Omega}{}^+$ partial-rate asymmetry, for which there is no previous measurement.

By angular-momentum conservation, in the decay of a spin-1/2 hyperon to a
spin-1/2 baryon plus a pion, the final state must be either $S$-wave or $P$-wave.\footnote{A similar argument holds for a spin-3/2 hyperon, but involving $P$ and $D$ waves.}
Interference between the $S$- and $P$-wave decay
amplitudes causes parity violation, described by Lee and
Yang\cite{Lee-Yang} in terms of two independent parameters $\alpha$ and
$\beta$, proportional to the real and imaginary parts
(respectively) of the interference term. Hyperon {\em CP}-violation signatures include differences in
$|\alpha|$ or $|\beta|$ between a hyperon decay and its {\em CP}-conjugate
antihyperon decay, as well as particle--antiparticle  decay partial-width differences between a mode and its {\em CP} conjugate.\cite{ACP,Donoghue-etal} Precision angular-distribution asymmetry measurement requires accurate knowledge of the relative polarizations of the
initial hyperons and antihyperons.

\subsubsection{Angular-distribution asymmetries}

Table~\ref{tab:HCP} summarizes the experimental situation.  
The first three experiments cited studied
$\Lambda$ decay only,\cite{R608}$^-$\cite{PS185} setting limits on the 
{\em CP}-asymmetry parameter\cite{ACP,Donoghue-etal}
\begin{eqnarray}
A_{\Lambda}\equiv \frac{\alpha_{\Lambda}+
\overline{\alpha}_\Lambda}{\alpha_{\Lambda}-\overline{\alpha}_\Lambda}\,,
\end{eqnarray}
where $\alpha_\Lambda$ ($\overline{\alpha}_\Lambda$) characterizes the
$\Lambda$ ($\overline{\Lambda}$) decay to  (anti)proton plus charged pion. If {\em CP} is a good symmetry in hyperon decay, $\alpha_\Lambda =
-\overline{\alpha}_\Lambda$. 

Fermilab fixed-target experiment E756\cite{E756} and CLEO\cite{CLEO} used the decay of charged $\Xi$ hyperons to produce polarized $\Lambda$'s, in whose subsequent decay the
slope of the (anti)proton angular distribution in the ``helicity" frame 
measures the product of $\alpha_\Xi$ and $\alpha_\Lambda$. If {\em
CP} is a good symmetry in hyperon decay this product should be identical for $\Xi^-$ and
$\overline{\Xi}{}^+$ events. The {\em CP}-asymmetry parameter measured is thus 
\begin{eqnarray}
A_{\Xi\Lambda}\equiv \frac{\alpha_{\Xi}\alpha_{\Lambda}-
\overline{\alpha}_\Xi\overline{\alpha}_\Lambda}{\alpha_{\Xi}\alpha_{\Lambda}+
\overline{\alpha}_\Xi\overline{\alpha}_\Lambda}
\approx A_\Xi + A_\Lambda\,.
\end{eqnarray}
The power of this technique derives from the relatively large $|\alpha|$ value for the
$\Xi^-\to\Lambda\pi^-$ decay ($\alpha_\Xi=-0.458\pm0.012$).\cite{PDG2006} 
A further advantage in the fixed-target case is that within a given
${}^{{}^(\!}\overline{\Xi}{}^{{}^)}$ momentum bin the acceptances and
efficiencies for $\Xi^-$
and $\overline{\Xi}{}^+$ decays are very similar, since the switch from detecting
$\Xi$ to detecting $\overline{\Xi}$ is made by reversing the polarities of the
magnets, making the spatial distributions of decay products across the detector
apertures almost identical for $\Xi$ and for $\overline{\Xi}$. 
(There are still residual systematic uncertainties
arising from the differing momentum dependences of the $\Xi$ and $\overline{\Xi}$ cross sections and  of the cross sections for the $p$ and $\overline{p}$ and
$\pi^+$ and $\pi^-$ to interact in the material of the spectrometer.)

Subsequent to E756, this technique was used in the ``HyperCP"
experiment (Fermilab E871),\cite{Holmstrom,Burnstein} which ran during 1996--99 and has set  the world's best limits on hyperon {\em CP} violation, based so far on about 5\% of the recorded ${}^{{}^(\!}\overline{\Xi}{}^{{}^)}{}^\mp\to{}^{{}^(\!}\overline{\Lambda}{}^{{}^)}\pi^\mp$ data sample.  (The systematics of the full data sample is still under study.) Like E756, HyperCP used a
secondary charged beam produced by 800\,GeV primary protons interacting in a metal
target. The secondary beam was  momentum- and sign-selected by means of a
curved 
collimator installed within a 6-m-long dipole magnet. Particle trajectories were
measured downstream of a 13-m-long (evacuated) decay region. HyperCP recorded the
world's largest samples of hyperon and antihyperon decays, including $2.0 \times 10^9$ and $0.46 \times 10^9$  $\Xi^-$ and $\overline{\Xi}{}^+$ events, respectively.   
When the analysis is complete, 
these should determine $A_{\Xi\Lambda}$ with a statistical uncertainty 
\begin{equation}
\delta A = \frac{1}{2\alpha_{\Xi}\alpha_{\Lambda}}
\sqrt{\frac{3}{N_{\Xi^-}}+\frac{3}{N_{\overline{\Xi}{}^+}}} \stackrel{<}{_\sim} 2\times10^{-4}\,.
\label{eq:ACP}
\end{equation} 
The standard model predicts\cite{ACP} this
asymmetry to be of order $10^{-5}$. 
Thus any significant effect seen in HyperCP will be evidence for new sources of {\em CP}
violation in the baryon sector. (A number of standard-model extensions, e.g., nonminimal SUSY, predict effects as large as ${\cal O}(10^{-3})$.\cite{non-SM}) Such an observation could be of relevance to the mysterious mechanism that gave rise to the  cosmic baryon asymmetry. 

HyperCP has also set the world's first limit on {\em CP} violation in ${}^{{}^(}\overline{\Omega}{}^{{}^)}{}^\mp$ decay, using a sample of 5.46~(1.89)~million $\Omega^-\to\Lambda K^-$  $({\overline\Omega}{}^+\to{\overline\Lambda} K^+)$ events.\cite{Lu-CP} Here, as shown by HyperCP,\cite{Chen,Lu} parity is only slightly violated: $\alpha=(1.75\pm0.24)\times10^{-2}$.\cite{PDG2006} Hence the measured magnitude and uncertainty of the  asymmetry parameter $A_{\Omega\Lambda}$ (inversely proportional to $\alpha$ as in Eq.~\ref{eq:ACP}) are rather large: $[-0.4\pm9.1\,(\rm stat)\pm8.5\,(syst)]\times10^{-2}$.\cite{Lu-CP} This asymmetry is predicted to be $\le4\times10^{-5}$ in the standard model but can be as large as $8\times10^{-3}$ if new physics contributes.\cite{Tandean}

\begin{table}[h]
\tbl {Summary of experimental limits on {\em CP} violation in hyperon decay; the hyperons studied are indicated by $^*$, $^\dagger$, and $^\ddag$.}
{\begin{tabular}{@{}lccccc@{}} \toprule
Exp't & Facility & Year & Ref. & Modes & $^*A_\Lambda\,/\,^\dagger A_{\Xi\Lambda}\,/\,^\ddag A_{\Omega\Lambda}$ \\
\colrule 
R608 & ISR & 1985 & \refcite{R608} & $pp\to\Lambda X, pp\to\overline{\Lambda} X$ &
$-0.02\pm0.14^*$ \\
DM2 & Orsay & 1988 &  \refcite{DM2} & $e^+e^- \to J/\psi \to \Lambda\overline{\Lambda}$ & 
$0.01\pm0.10^*$
\\
PS185 & LEAR & 1997 & \refcite{PS185} & $\overline{p}p\to\overline{\Lambda}\Lambda$ & 
$0.006\pm0.015^*$ \\
& & & &
$e^+e^-\to\Xi^- X, \Xi^-\to\Lambda\pi^-,$ &\\
\raisebox{1.5ex}[0pt]{CLEO} & \raisebox{1.5ex}[0pt]{CESR} &\raisebox{1.5ex}[0pt]{2000} &
\raisebox{1.5ex}[0pt]{\refcite{CLEO}} & $e^+e^-\to\overline{\Xi}{}^+ X, \overline{\Xi}{}^+\to\overline{\Lambda}\pi^+$ & 
\raisebox{1.5ex}[0pt]{$-0.057\pm0.064\pm
0.039^\dagger$}\\ 
&  & & &
$pN\to\Xi^- X, \Xi^-\to\Lambda\pi^-$, &  \\
\raisebox{1.5ex}[0pt]{E756} &\raisebox{1.5ex}[0pt]{FNAL} & \raisebox{1.5ex}[0pt]{2000} &
\raisebox{1.5ex}[0pt]{\refcite{E756}} & $pN\to\overline{\Xi}{}^+ X, \overline{\Xi}{}^+\to\overline{\Lambda}\pi^+$ & 
\raisebox{1.5ex}[0pt]{$0.012
\pm0.014^\dagger$} \\
&  & & &
$pN\to\Xi^- X, \Xi^-\to\Lambda\pi^-$, &  \\
\raisebox{1.5ex}[0pt]{HyperCP} &
\raisebox{1.5ex}[0pt]{FNAL} & \raisebox{1.5ex}[0pt]{2004} &  \raisebox{1.5ex}[0pt]{\refcite{Holmstrom}} &
$pN\to\overline{\Xi}{}^+ X, \overline{\Xi}{}^+\to\overline{\Lambda}\pi^+$ & 
\raisebox{1.5ex}[0pt]{$(0.0
\pm6.7)\times10^{-4}{}^{\,\dagger,\S}$} \\
&  & & &
$pN\to\Omega^- X, \Omega^-\to\Lambda K^-$, &  \\
\raisebox{1.5ex}[0pt]{HyperCP} &
\raisebox{1.5ex}[0pt]{FNAL} & \raisebox{1.5ex}[0pt]{2006} &  \raisebox{1.5ex}[0pt]{\refcite{Lu-CP}} &
$pN\to\overline{\Omega}{}^+ X, \overline{\Omega}{}^+\to\overline{\Lambda} K^+$ & 
\raisebox{1.5ex}[0pt]{$-0.004\pm 0.12^{\,\ddag}$} \\
\botrule 
\end{tabular}
\label{tab:HCP}}
\footnotesize$^\S$ Based on $\approx$5\% of the HyperCP data sample; analysis of the full sample is still in progress.
\end{table}

\subsubsection{Partial-rate asymmetries}

While {\em CPT} symmetry requires identical lifetimes for particle and antiparticle, partial-rate asymmetries violate only {\em CP}. For most hyperon decays, these are expected to be undetectably small.\cite{Valencia2000} However, for the decays $\Omega^- \to \Lambda K^-$ and $\Omega^- \to \Xi^0\pi^-$, the particle/antiparticle partial-rate asymmetries could be as large as $2\times10^{-5}$ in the standard model and one to two orders of magnitude larger if non-SM contributions dominate.\cite{Tandean-Valencia,Tandean} The quantities to be measured are
\begin{eqnarray}\nonumber
\Delta_{\Lambda K} &\equiv&\frac{\Gamma(\Omega^-\to\Lambda K^-)-\Gamma({\overline \Omega}{}^+\to{\overline\Lambda}K^+)}{\Gamma(\Omega^-\to\Lambda K^-)+\Gamma({\overline \Omega}{}^+\to{\overline\Lambda}K^+)}\\
&\approx&\frac{1}{2\Gamma}(\Gamma-{\overline \Gamma})
\approx 0.5\,(1-N/{\overline N})
\end{eqnarray}
(and similarly for $\Delta_{\Xi \pi}$),
where in the last step we have assumed nearly equal numbers ($N$) of $\Omega$ and (${\overline N}$) of ${\overline \Omega}$ events, as would be the case in ${\overline p}p$ annihilation. Sensitivity at the $10^{-4}$ level then requires ${\cal O}(10^7)$ reconstructed events. 
Measuring such a small branching-ratio difference reliably will require the clean exclusive ${\overline\Omega}{}^+\Omega^-$ event sample produced less than a $\pi^0$ mass above threshold, or $4.938<p_{\overline p}< 5.437$\,GeV/$c$. 

\subsubsection{Hyperon sensitivity estimates}

There have been a number of measurements of hyperon production by low-energy antiprotons. Johansson {\it et al.}\cite{Johansson} report cross sections measured by PS185 at LEAR, but the maximum LEAR $\overline p$ momentum (2\,GeV/$c$) was insufficient to produce $\Xi$'s or $\Omega$'s. Chien {\it et al.}\cite{Chien} report measurements  of a variety of hyperon final states performed with the BNL 80-inch liquid-hydrogen bubble chamber in a 6.935\,``BeV/$c$" electrostatically separated antiproton beam at the AGS; Baltay {\it et al.}\cite{Baltay} summarize data taken at lower momenta. In 80,000 pictures Chien {\it et al.}\ observed some 1,868 hyperon or antihyperon events, corresponding to a total hyperon-production cross section of $1.310\pm0.105$\,mb.\cite{Chien} The corresponding cross section measured at 3.7\,GeV/$c$ was $720\pm30\,\mu$b, and $438\pm52\,\mu$b at 3.25\,GeV/$c$.\cite{Baltay} The inclusive hyperon-production cross section at 5.4\,GeV/$c$ is thus about 1\,mb.  At $2\times10^{32}\,\rm cm^{-2}s^{-1}$ this amounts to some $2\times10^5$ hyperon events produced per second, or $2\times10^{12}$ per year. (Experience suggests that a data-acquisition system that can cope with such a high event rate is both feasible and reasonable in cost. For example, the ${\overline p}p$ interaction rate is comparable to that in BTeV, yet the charged-particle multiplicity per event is only $\approx$\,1/10 as large.)

To estimate the exclusive ${\overline p}p\to{\overline\Omega}\Omega$ cross section requires some extrapolation, since it has yet to be measured (moreover, even for ${\overline p}p\to{\overline \Xi}{}^+\Xi^-$ only a few events have been seen). A rule of thumb is that each strange quark ``costs" between one and two orders of magnitude in cross section, reflecting the effect of the strange-quark mass on the hadronization process.  This is borne out e.g.\ by HyperCP, in which $2.1\times10^9$ $\Xi^-\to\Lambda \pi^-$and $1.5\times10^7$ $\Omega^-\to\Lambda K^-$ decays were reconstructed;\cite{Burnstein} given the 160\,GeV/$c$ hyperon momentum and 6.3\,m distance from  HyperCP target to decay pipe, this corresponds to $\approx$\,30 $\Xi^-$'s per $\Omega^-$ produced at the target. A similar ratio is observed in HERA-$B$.\cite{Britsch} In exclusive ${\overline p}p\to{\overline Y}Y$ production (where $Y$ signifies a hyperon) there may be additional effects, since as one proceeds from $\Lambda$ to $\Xi$ to $\Omega$ fewer and fewer valence quarks are in common between the initial and final states. Nevertheless, the  cross section for ${\overline \Xi}{}^+\Xi^-$ somewhat above threshold ($p_{\overline p}\approx3.5\,$GeV/$c$) is $\approx$\,2\,$\mu$b,\cite{Hamann,Baltay,HERAG} or about 1/30 of the corresponding cross section for ${\overline \Lambda}\Lambda$.
Thus the $\approx$\,65\,$\mu$b cross section measured for ${\overline p}p\to{\overline \Lambda}\Lambda$ at $p_{\overline p}=1.642$\,GeV/$c$ at LEAR\cite{Johansson} implies $\sigma({\overline p}p\to{\overline \Omega}\Omega)\sim60$\,nb at 5.4\,GeV/$c$. 

For purposes of discussion we take this as the exclusive production cross section.\footnote{This estimate will be testable in the upgraded MIPP experiment.\protect\cite{MIPP-upgrade}} At  $2.0\times10^{32}\,{\rm cm}^{-2}{\rm s}^{-1}$ luminosity, some $1.2\times10^8$ ${\overline \Omega}\Omega$ events are then produced in a nominal 1-year run ($1.0\times10^7$\,s). Assuming 50\% acceptance times efficiency  (comparable to that for $\chi_c$ events in E760), we estimate ${}^{{}^(\!}\overline {N}{}^{{}^)}_{\Xi\pi}=1.4\times10^7$ events each in $\Omega^-\to\Xi^0\pi^-$ and ${\overline \Omega}{}^+\to{\overline\Xi}{}^0\pi^+$, and ${}^{{}^(\!}\overline {N}{}^{{}^)}_{\Lambda K}=4.1\times10^7$ events each in $\Omega^- \to \Lambda K^-$  and ${\overline\Omega}{}^+ \to {\overline\Lambda} K^+$, implying the  partial-rate-asymmetry statistical sensitivities
\begin{eqnarray}
\delta\Delta_{\Xi\pi}\approx\frac{0.5}{\sqrt{N_{\Xi\pi}}}\approx1.3\times10^{-4}\,,~~
\delta\Delta_{\Lambda K}\approx\frac{0.5}{\sqrt{N_{\Lambda K}}}\approx7.8\times10^{-5}
\,.
\end{eqnarray}
Tandean and Valencia\cite{Tandean-Valencia} have estimated $\Delta_{\Xi\pi}\approx2\times10^{-5}$ in the standard model but possibly an order of magnitude larger with new-physics contributions.
Tandean\cite{Tandean} has estimated $\Delta_{\Lambda K}$ to be $\le1\times10^{-5}$ in the standard model but possibly as large as $1\times10^{-3}$ if new physics contributes. (The large sensitivity of $\Delta_{\Lambda K}$ to new physics in this analysis  arises from chromomagnetic penguin operators and final-state interactions via $\Omega\to\Xi\pi\to\Lambda K$.\cite{Tandean}\,\footnote{Large final-state interactions  should also affect $\Delta_{\Xi\pi}$ but were not included in that prediction.\protect\cite{Tandean-Valencia,Tandean-private}}) It is worth noting that these potentially large asymmetries arise from parity-conserving interactions and hence are limited by constraints from  $\epsilon_K$;\cite{Tandean-Valencia,Tandean} they are independent of $A_\Lambda$ and $A_\Xi$, which arise from the interference of parity-violating and parity-conserving processes.\cite{Tandean-private} 

Of course, the experimental sensitivities will include systematic components whose estimation will require careful and detailed simulation studies yet to be done. Nevertheless, the potential power of the technique is apparent: the experiment discussed here may be capable of observing the effects of new physics in Omega {\em CP} violation via partial-rate asymmetries, and it will  represent a substantial improvement over current sensitivity to Omega angular-distribution asymmetries.

\subsection{Study of FCNC hyperon decays}

\begin{figure}
\centerline{
\includegraphics[width=0.42\linewidth]{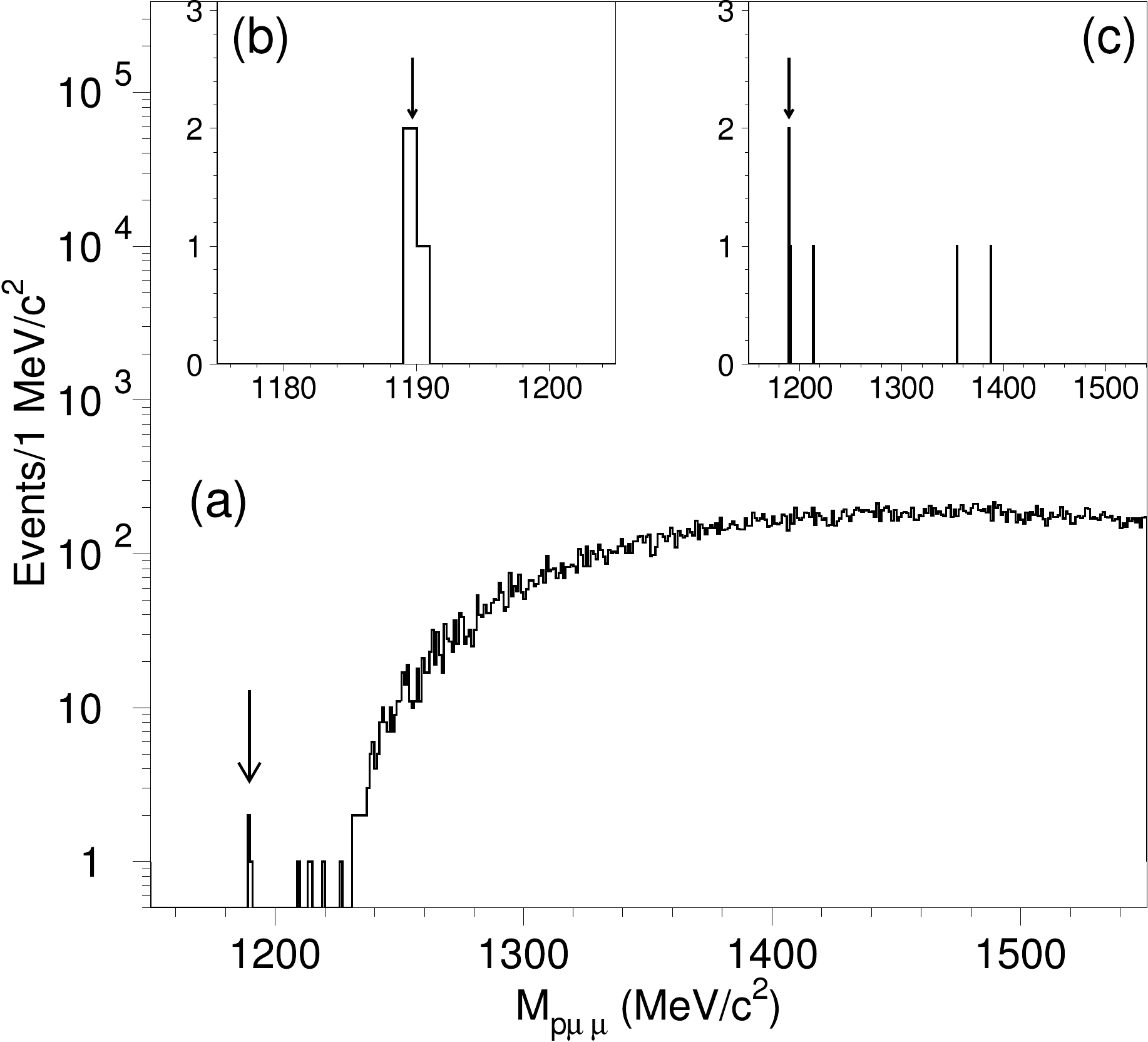}\hspace{0.1in}\includegraphics[width=0.5\linewidth]{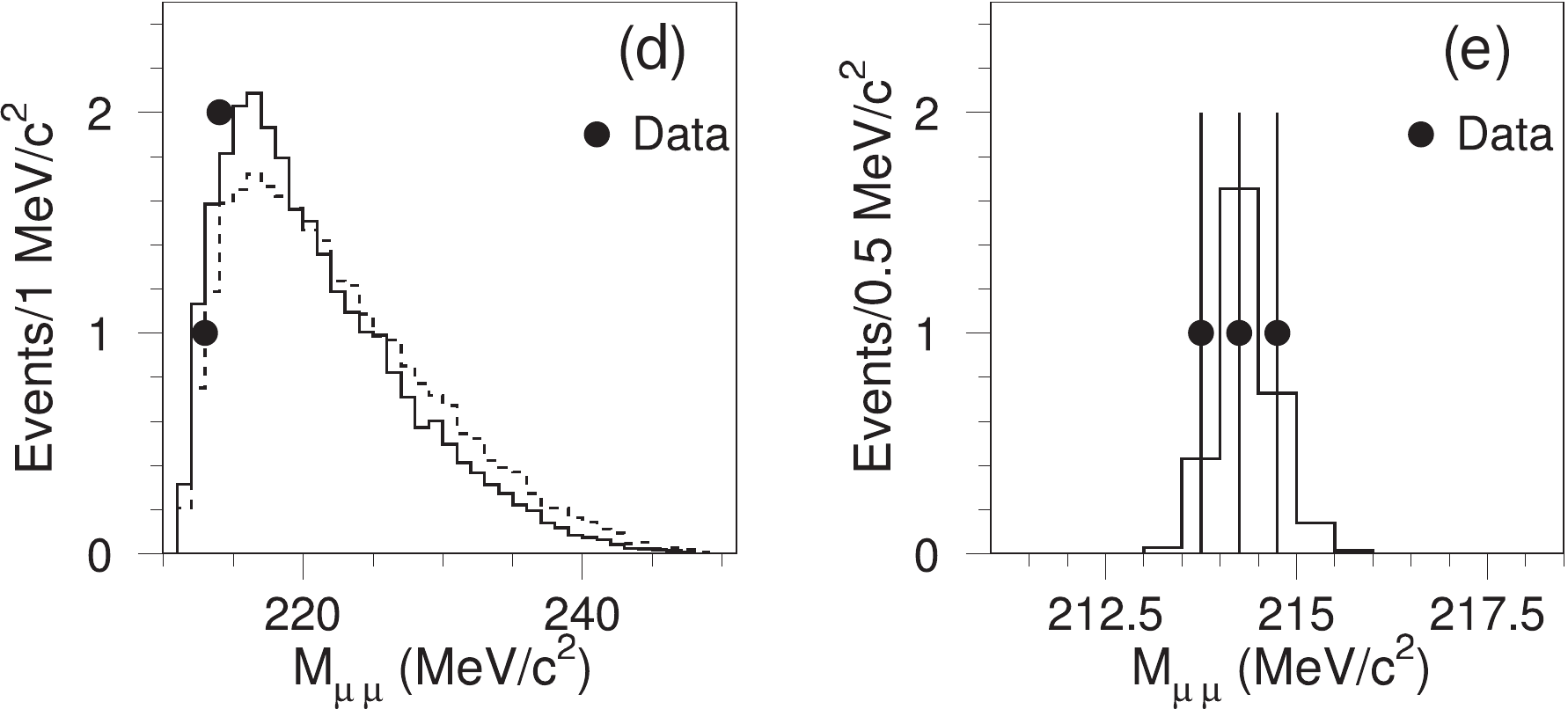}}
\caption{Mass spectra for candidate  single-vertex $p\mu^+\mu^-$  events in HyperCP positive-beam data sample: (a) wide mass range (semilog scale); (b) narrow range around $\Sigma^+$ mass; (c) after application of additional cuts as described in Ref.~\protect\refcite{Park-Sigpmumu} (arrows indicate mass of $\Sigma^+$);  dimuon mass spectrum of the
candidate events compared with Monte Carlo spectrum assuming (d) standard-model virtual-photon form factor (solid) or isotropic decay (dashed), or (e) decay via a narrow resonance $X^0$.}
\label{fig:sigpmumu}
\label{fig:mumu}
\end{figure}

Behind its charged-particle spectrometer, HyperCP had muon detectors for  rare-decay studies.\cite{Burnstein,Park-Sigpmumu}  Using them HyperCP has observed\cite{Park-Sigpmumu}  the rarest hyperon decay ever, $\Sigma^+\to p\mu^+\mu^-$. Surprisingly (Fig.~\ref{fig:sigpmumu}),  the 3 observed events are consistent with a two-body decay, $\Sigma^+\to p X^0,\,X^0\to\mu^+\mu^-$, with $X^0$ mass $m_{X^0}=214.3\pm0.5\,$MeV/$c^2$. This interpretation is of course not definitive, with the confidence level for the form-factor decay spectrum of Fig.~\ref{fig:mumu}d  estimated at 0.8\%. The measured branching ratio is $[3.1\pm2.4\, (\rm stat)\pm1.5\, (syst)]\times 10^{-8}$ assuming two-body, or $[8.6^{+6.6}_{-5.4}\, (\rm stat)\pm  5.5\,(syst)] \times10^{-8}$ assuming three-body $\Sigma^+$ decay.

This result is intriguing in view of Gorbunov's proposal\cite{Gorbunov} that  certain nonminimal supersymmetric models include a pair of ``sgoldstinos" (supersymmetric partners of Goldstone fermions), which can be scalar or pseudoscalar and  low in mass. A light scalar particle coupling to hadrons and  muon pairs at the required level is ruled out by its nonobservation in kaon decays; however, a pseudoscalar sgoldstino with $\approx$\,214\,MeV/$c^2$ mass would be consistent with all available data.\cite{He-etal-Sigpmumu}$^-$\cite{Gengetal} Alternatively, He, Tandean, and Valencia suggest\cite{He-Tandean-Valencia} the $X^0$ is the light pseudoscalar Higgs boson ($A^0_1$) in the next-to-minimal supersymmetric standard model.

Studying this with exclusive ${\overline \Sigma}{}^-\Sigma^+$ events just above threshold would require ${\overline p}$ momentum (see Table~\ref{tab:thresh}) well below that previously achieved by deceleration in the Antiproton Accumulator, as well as very high luminosity to access the ${\cal O}(10^{-8})$ branching ratio. An experimentally less challenging but equally interesting objective is the corresponding FCNC decay of the $\Omega^-$, with  ${\cal O}(10^{-6})$ predicted branching ratio\cite{He-etal-Sigpmumu} if the $X^0$ is real.\footnote{The standard-model prediction is\protect\cite{Safadi-Singer} ${\cal B}( \Omega^-\to\Xi^-\mu^+\mu^-)=6.6\times10^{-8}$.}
(The larger  branching ratio reflects the additional phase space available compared to that in $\Sigma^+\to p\mu^+\mu^-$.) As above, assuming $2\times10^{32}$ luminosity and 50\% acceptance times efficiency, 120 or 44 events are predicted in the two cases (pseudoscalar or axial-vector $X^0$) that appear to be viable:\cite{He-etal-Sigpmumu,Deshpande-Eilam-Jiang}
\begin{eqnarray}
{\cal B}(\Omega^-\to\Xi^- X_P\to \Xi^-\mu^+\mu^-)&=& (2.0^{+1.6}_{-1.2}\pm1.0)\times10^{-6}\,,\\ 
{\cal B}(\Omega^-\to\Xi^- X_A\to \Xi^-\mu^+\mu^-)&=&(0.73^{+0.56}_{-0.45}\pm 0.35)\times10^{-6}\,.
\end{eqnarray}
Given the large inclusive hyperon rates at $\sqrt{s}\approx 3.5\,$GeV, sufficient sensitivity might also be available at that setting to confirm the HyperCP $\Sigma^+\to p\mu^+\mu^-$ results. Alternatively, it is possible that a dedicated run just above $\overline{\Sigma}{}^-\Sigma^+$ threshold may have competitive sensitivity; evaluating this will require a detailed simulation study.

\subsection{Additional physics}\label{addl-phys}

Besides the $X(3872)$, the experiment would be competitive for the charmonium and related states mentioned in Sec.~\ref{physics-list}. The large hyperon samples could enable precise measurement of hyperon semileptonic and other rare decays. The APEX experiment\cite{APEX} vacuum tank and pumping system could be reinstalled, enabling substantially increased sensitivity for $\overline p$ lifetime and decay modes. 
There is interest in decelerating further (e.g., at the ends of stores) for trapped-antiproton and antihydrogen experiments.\cite{Holzscheiter,Phillips}  This capability could make Fermilab the premiere facility for such research. The $\overline p$ intensity available at Fermilab could enable  studies not feasible at the AD, such as a measurement of the gravitational force on antimatter.\cite{Phillips}  A complementary approach is the study of antihydrogen atoms in flight,\cite{Blanford} which may overcome some of the difficulties encountered in the trapping experiments.

The PANDA TPR\cite{PANDA-TPR} claims competitive sensitivity for open charm, estimating the rate of $D$-pair production at about 100/s for $\sqrt{s}$ near the $\psi(4040)$. This could lead to a sample of $\sim10^9$\,events/year produced and  $\sim10^8$/year reconstructed, roughly an order of magnitude beyond the statistics accumulated by the B Factories so far. Whether this sensitivity can be realized in practice will depend on details of trigger and analysis efficiency whose estimation will require detailed simulation studies. Nevertheless, there does appear to be the potential for  competitive measurements, e.g., of $D^0$ mixing and possible {\em CP} violation in charm decay.

The bottomonium system has not benefited from $\overline{p}p$ formation studies but is potentially accessible if the Antiproton Accumulator (or perhaps a new replacement storage ring) can be configured for colliding beams. The $\overline{p}p$ widths of bottomonium states are unknown. If they can be shown to be sufficiently large, $\overline{p}p$ formation could lead to the discovery of bottomonium singlet states, which have so far eluded observation, as well as precise measurements of the many states already observed.

\section{A New Experiment}

We see two approaches to implementing low-cost apparatus to perform the measurements here described:\cite{pbar-LoI} one based on existing equipment from E835, and the other  on the D\O\ superconducting solenoid (available once the Tevatron Collider program ends). 
Should sufficient resources be available, a new spectrometer, free of constraints from existing apparatus, may give better performance than either of these. The possibility of building a new storage ring has also been mentioned. We hope to study these options in detail in the coming months. 
\section*{Acknowledgments}

This work was supported by the US Dept.\ of Energy under grant DEFG02-94ER40840. The author thanks all of his pbar collaborators and especially D. Christian, K. Gollwitzer, G. Jackson, R. Mussa, C. Patrignani, S. Pordes, J. Rosen, and J. Rosner for useful and stimulating conversations.

\end{document}